\documentclass[apj]{emulateapj}
\submitted{}
\usepackage{lscape}
\usepackage{apjfonts}
\usepackage{graphicx}
\def\kms{{\rm km}\,{\rm s}^{-1}}
\def\hbn{{\hfil\break\noindent}}

\begin{document}
\title{Investigation of the Contamination of the Gould (2003) Halo Sample}

\author{
Andrew~Gould\altaffilmark{1}
}
\altaffiltext{1}
{Department of Astronomy, Ohio State University,
140 W.\ 18th Ave., Columbus, OH 43210, USA; 
gould@astronomy.ohio-state.edu}
\begin{abstract}

 A recent astroph posting argued
that the \citet{gould03a} halo sample
is substantially contaminated with thick-disk stars, which would
then ``wash out'' any signature of granularity in the halo
velocity distribution due to streams.  If correct, this would imply that
the limits placed by \citet{gould03b} on streams are not valid.
Here I investigate such contamination using six different indicators,
\hbn 1) morphology of the underlying reduced proper motion diagram used
to select halo stars,
\hbn 2) comparison of kinematic and parallax-based distance scales
\hbn 3) comparison of derived halo parameters for the \citet{gould03a}
sample with other determinations
\hbn 4) a precision color-color diagram for a random subsample 
\hbn 5) the 3-dimensional velocity distribution of a random subsample
\hbn 6) metallicity distribution versus kinematic cuts on a random subsample
\hbn I estimate that the contamination is of order 2 percent.
Thus, the upper limits on the density of nearby streams derived by
\citet{gould03b} remain valid.
In particular, at 95\% confidence, no more than 5\% of local halo
stars (within about $300\,$pc) are in any one coherent stream.
 Determining whether or
not this local measurement is consistent with CDM {\it remains} an
outstanding question.
\end{abstract}


\section{Introduction
\label{sec:intro}}
A critical test of the current picture of hierarchical structure
formation, is measuring the "granularity" or substructure within the
halo as a function of Galactocentric distance.  \citet{gould03b} 
determined that,
contrary to naive expectation within this framework, the local
Galactic halo is remarkably smooth, with only $\sim 5\%$ possible contribution
from substructures.  However, this conclusion is only as good as the
underlying \citet{gould03a} sample: significant contamination by
thick-disk stars would tend
to ``wash out'' any substructure signal \citep{kepley07}. 
Prompted by this concern, I demonstrate below through 
six different tests, that this sample is not in fact seriously 
contaminated and therefore that
the current (still valid) limit of $\sim 5\%$ 
substructure must be understood by current models of Galaxy formation.

\section{Morphology of the rNLTT Reduced Proper Motion Diagram
\label{sec:morphology}}

The \citet{gould03a} halo sample is derived from the revised New Luyten
Two-Tenths (rNLTT) catalog of \citet{gs03} and \citet{sg03}.
The argument made by \citet{kepley07} for contamination of this
sample is that \citet{ryan91} found such contamination among the halo
candidates that they had extracted from the underlying \citet{luy} NLTT 
catalog.

\begin{figure}
\includegraphics[angle=0,width=3.25in]{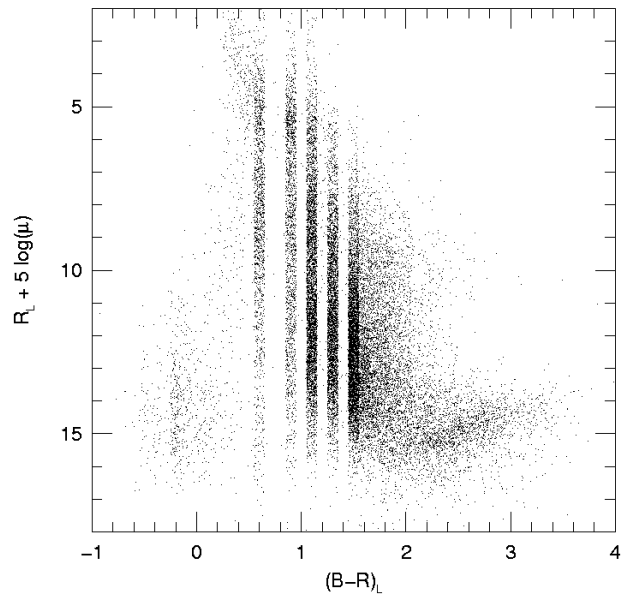}
\caption{Reduced proper motion diagram for the original NLTT catalog,
i.e., using the original proper motions and, more importantly,
the original photographic ($B$, $R$) photometry.  (Original $B-R$ was
sometimes given to only 1 decimal place: for these, small random
numbers have been added to the color to permit display).  From 
\citet{sg02}.}
\label{fig:rpml}
\end{figure}

\begin{figure}
\includegraphics[angle=0,width=3.25in]{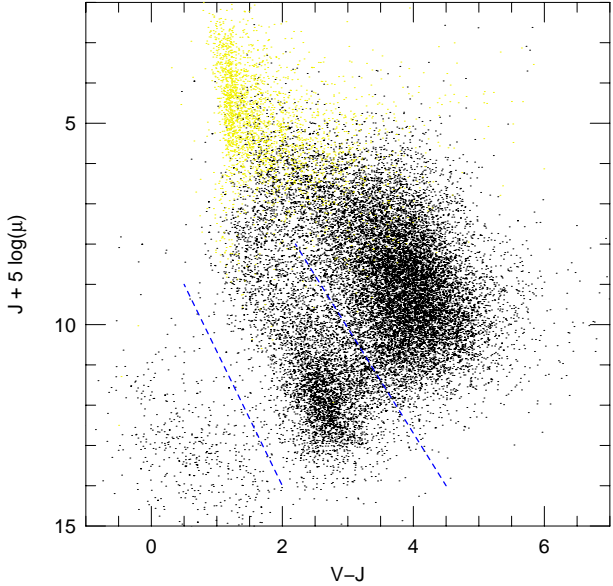}
\caption{Reduced proper motion diagram for {\it revised} rNLTT catalog
\citep{gs03,sg03} i.e., using the new proper motions and, more importantly,
new 2MASS $J$ photometry to enable a much broader-baseline and more accurate
$V-J$ color.  From \citet{sg02}.}
\label{fig:rpmj}
\end{figure}

Figures \ref{fig:rpml} and \ref{fig:rpmj} (both taken from \citealt{sg02})
show the reduced proper motion (RPM) diagrams for the 
NLTT and rNLTT, respectively.  
Halo stars are clearly separated from disk/thick-disk
stars in rNLTT (used by \citealt{gould03a}), 
but not NLTT (used by \citealt{ryan91}) .

\begin{figure}
\includegraphics[angle=0,width=3.25in]{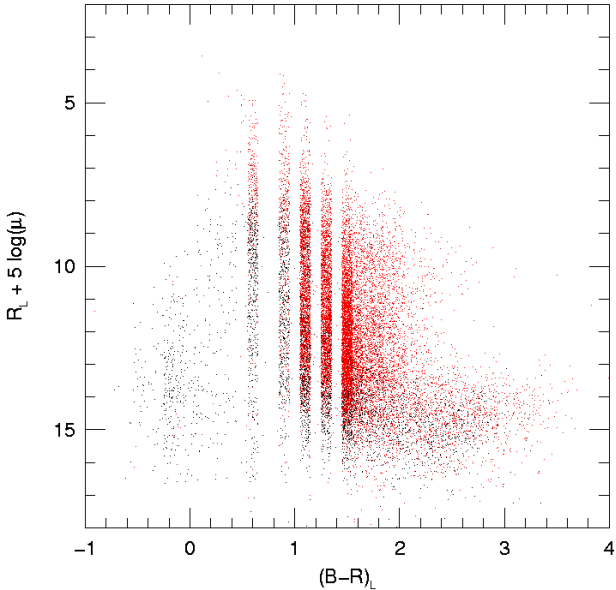}
\caption{Reduced proper motion diagram for the original NLTT catalog,
i.e., same as Fig.\ \ref{fig:rpml} restricted to rNLTT$\cap$NLTT,
but with main-sequence stars
(as determined from Fig.\ \ref{fig:rpmj}) shown in red.
From \citet{sg02}.}
\label{fig:rpmlms}
\end{figure}

\begin{figure}
\includegraphics[angle=0,width=3.25in]{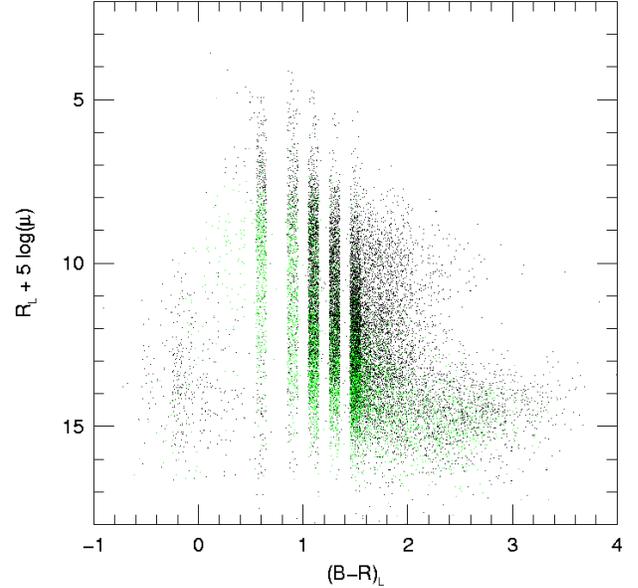}
\caption{Reduced proper motion diagram for the original NLTT catalog,
i.e., same as Fig.\ \ref{fig:rpml}restricted to rNLTT$\cap$NLTT, 
but with subdwarfs
(as determined from Fig.\ \ref{fig:rpmj}) shown in green.
By comparing this figure with Fig.\ \ref{fig:rpmlms}, it is clear that
it is impossible to design selection criteria that would recover
a large number of subdwarfs from the original NLTT
without substantial contamination by
main-sequence stars.  From \citet{sg02}.}
\label{fig:rpmlsd}
\end{figure}

Figures \ref{fig:rpmlms} and \ref{fig:rpmlsd}
further confirm the difficulty of extracting a clean sample
of halo stars from NLTT.  They show the distributions of main-sequence
stars and subdwarfs on the NLTT RPM.  Clearly, there is no way to
select subdwarfs from Luyten's original NLTT 
without substantial main-sequence contamination.

Subsequently, \citet{sg03} introduced a further 
refinement of the optical/infrared
RPM of \citet{sg02} by defining a discriminator $\eta$ that depends
on both RPM and Galactic latitude $b$:
\begin{equation}
\eta = V + 5\log\mu - 3.1(V-J) - 1.47|\sin b| - 2.73
\label{eqn:eta}
\end{equation}
where $\mu$ is the proper motion in arcsec per year.  They identified
halo stars as dominant in the range
\begin{equation}
0<\eta < 5.15\qquad {[\rm Salim\ \&\ Gould\ (2003):\ halo\ stars]}
\label{eqn:etaorig}
\end{equation}

\begin{figure}
\includegraphics[angle=0,width=3.25in]{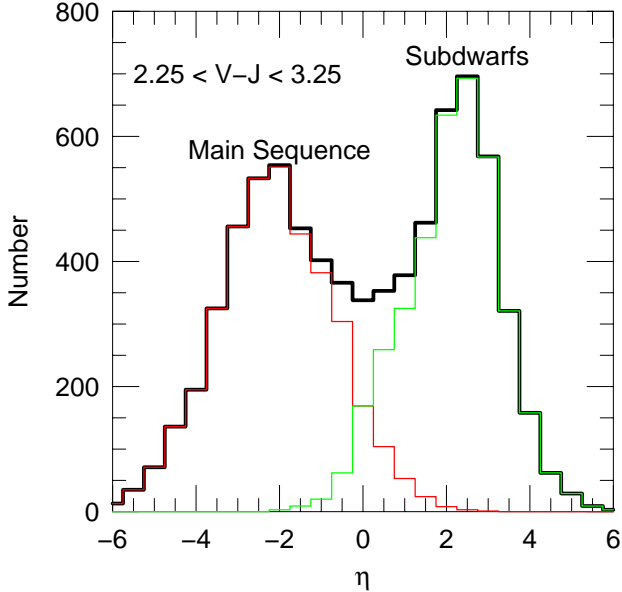}
\caption{Distribution of ``population discriminator'' $\eta$ for rNLTT
stars with $2.25<V-J<3.25$.  There are two clear peaks corresponding
to main-sequence stars (red) and subdwarfs (green).
In the overlap the bins have been divided by symmetrizing (see text).
Adapted from \citet{sg03}.}
\label{fig:eta}
\end{figure}

Figure \ref{fig:eta}, adapted from \citet{sg03} shows (in black)
the distribution of $\eta$ in the color range $2.25<V-J<3.25$.
This confirms quantitatively the visual impression from Figure \ref{fig:rpmj}
that there is a clear valley between the subdwarfs and the main-sequence
stars.  To further clarify the situation I have estimated the 
subdwarf/main-sequence breakdown in the region of overlap as follows
(green and red histograms in figure).  First, at the valley minimum
($\eta=0$) I assigned half the stars to each population.  Second,
for $\eta<0$, I estimated the falling profile of subdwarfs to be the
mirror image of their rising profile at high $\eta$.  Third,
for $\eta>0$, I estimated the falling profile of main-sequence stars to
be the mirror image their rising profile at low $\eta$.  

In order to obtain a pure sample of halo stars, \citet{gould03a}
further restricted the range of $\eta$ relative to 
equation (\ref{eqn:etaorig}),
using
\begin{equation}
1<\eta < 4.15\qquad {[\rm Gould\ (2003):\ secure\ halo\ stars]}
\label{eqn:etasecure}
\end{equation}
to avoid main-sequence stars at high $\eta$ and white dwarfs at low $\eta$.
 From Figure \ref{fig:eta}, one may estimate that this choice generates
roughly 2\% contamination.

While I will give several other independent arguments that the sample
is not seriously contaminated, this one is the strongest and most
quantitative.

\section{Comparison of Parallax and Kinematic Distance Scales
\label{sec:parallax}}

\begin{figure}
\includegraphics[angle=0,width=3.25in]{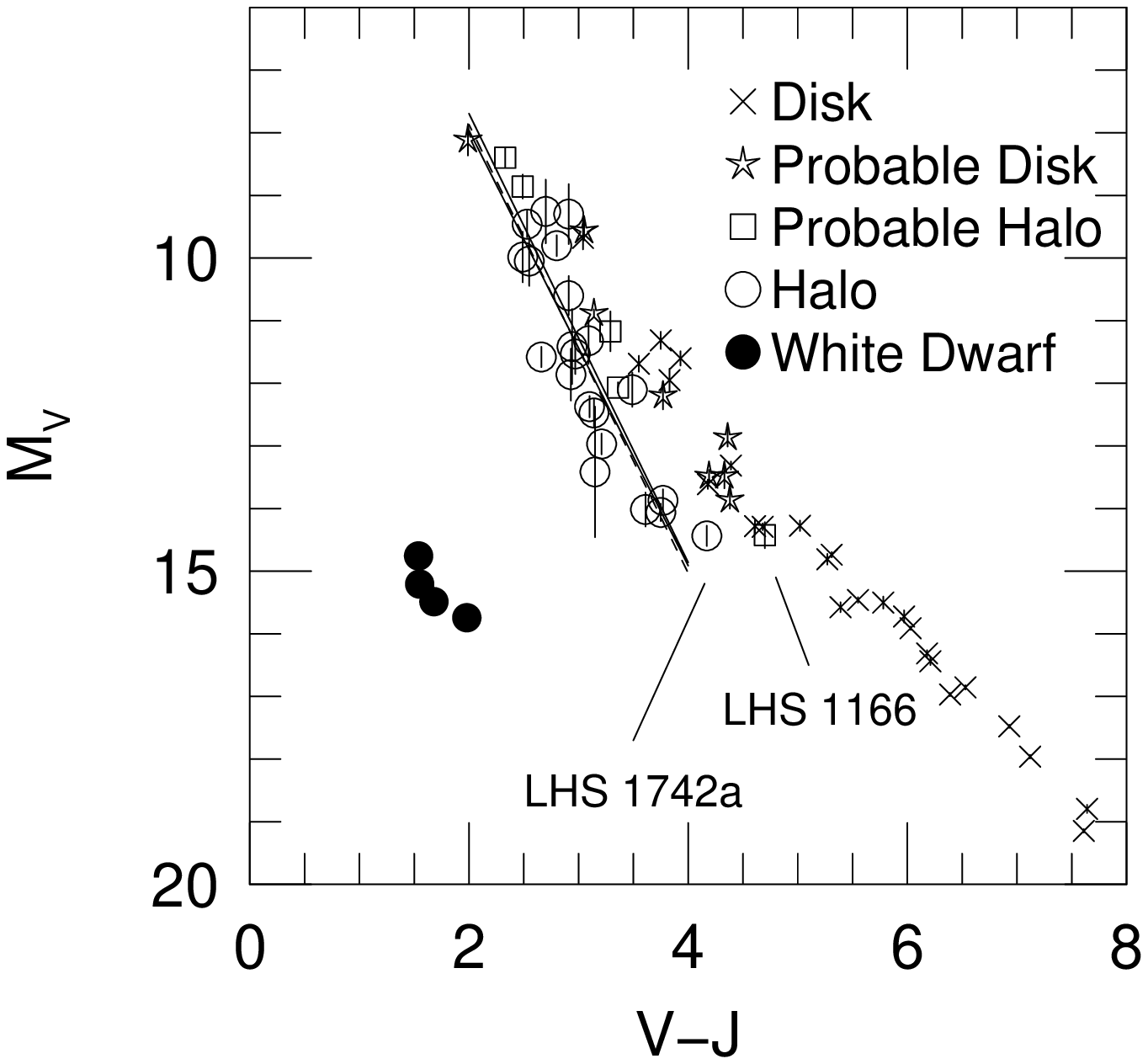}
\caption{CMD of stars with trigonometric parallaxes \citep{monet92,gizis97},
separated into various classes using the \citet{sg03} RPM-discriminator
$\eta$.  The solid line shows the best fit to the halo stars in the data
while the dashed line shows the fit based on kinematic analysis of
the \citet{gould03a} sample.  They are virtually identical. If there
were serious contamination by thick disk stars, the dashed line would
have been driven to brighter mags.  From \citet{gould03a}}
\label{fig:monet3}
\end{figure}

A second check comes by comparing the kinematic-based distance scale
derived by \citet{gould03a} with trigonometric parallaxes found in 
the literature.  To establish distances, \citet{gould03a} fit for
the two parameters of a linear color-magnitude relation, while enforcing
a mean motion of halo stars relative to Sun of $U_2 = -216.6\,\kms$,
so as to force agreement with the determination of this parameter
by \citet{gp98} based on halo RR Lyrae stars.  \citet{gould03a} then
compared this with the color-magnitude relation derived from
halo stars with trigonometric parallaxes from \citet{monet92} and 
\citet{gizis97}.  As can be seen from Figure \ref{fig:monet3}, these
relations are virtually identical.  If the sample were contaminated
with thick-disk stars (whose mean motion $U_2$ motion relative to the
Sun is of order 5 times smaller than halo stars) by even 10\%, then 
this would cause an error in the distance scale of $0.80\times 10\% = 8\%$,
yielding an offset between the parallax and kinematic relations of 
about 0.17 mag.  The offset is clearly much smaller than this.

\section{Comparison of Halo Kinematics
\label{sec:kinematics}}

It is not only the asymmetric drift of the halo that is correctly
reproduced by the \citet{gould03a} analysis, but also the velocity
dispersions.  \citet{gould03a} found dispersions in the radial,
rotation, and vertical directions of $168\pm 1$, $113\pm 2$, and $89\pm 2\ 
\kms$, respectively.  These values are quite compatible with other
determinations.  For example \citet{gp98} found $171\pm 10$, 
$99\pm 8$, and $90\pm 7$ from a much smaller sample of halo RR Lyrae
stars.  Serious contamination by the kinematically 
much cooler population of
thick-disk stars would have tended to drive down these dispersions
(assuming the distance scale remained in agreement with the parallax
stars of Fig.\ \ref{fig:monet3}).

\section{Precision Color-Color Diagram of a Random Subsample
\label{sec:color}}

\begin{figure}
\includegraphics[angle=0,width=3.25in]{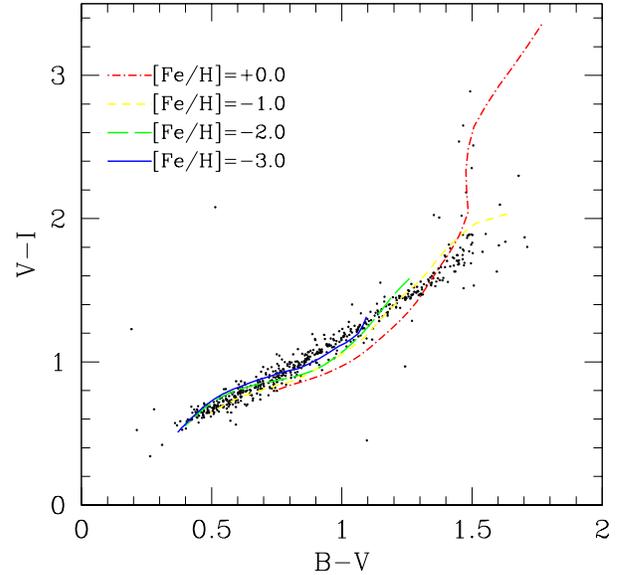}
\caption{$BVI$ color-color plot of 564 stars selected from rNLTT using
the same $1<\eta<4.15$ criterion applied by \citet{gould03a}.  The 
sample is metal poor and relatively homogeneous.  From \citet{marshall07}.}
\label{fig:marshall9}
\end{figure}

Marshall has obtained precise photometric data for 564 stars selected by
applying the same $1<\eta<4.15$ criterion used
by \citet{gould03a}.  Her Figure 9 (reproduced here as 
Fig.~\ref{fig:marshall9}) shows a very tight color-color relation for these
stars, consistent with a metal-poor population.  It is clear that the 
stars extracted by this criterion are not very heterogeneous, as
would have been expected were they seriously contaminated.

\section{3-D Velocity Distribution of a Random Subsample}
\label{sec:velocity}

\begin{figure}
\includegraphics[angle=0,width=6.5in]{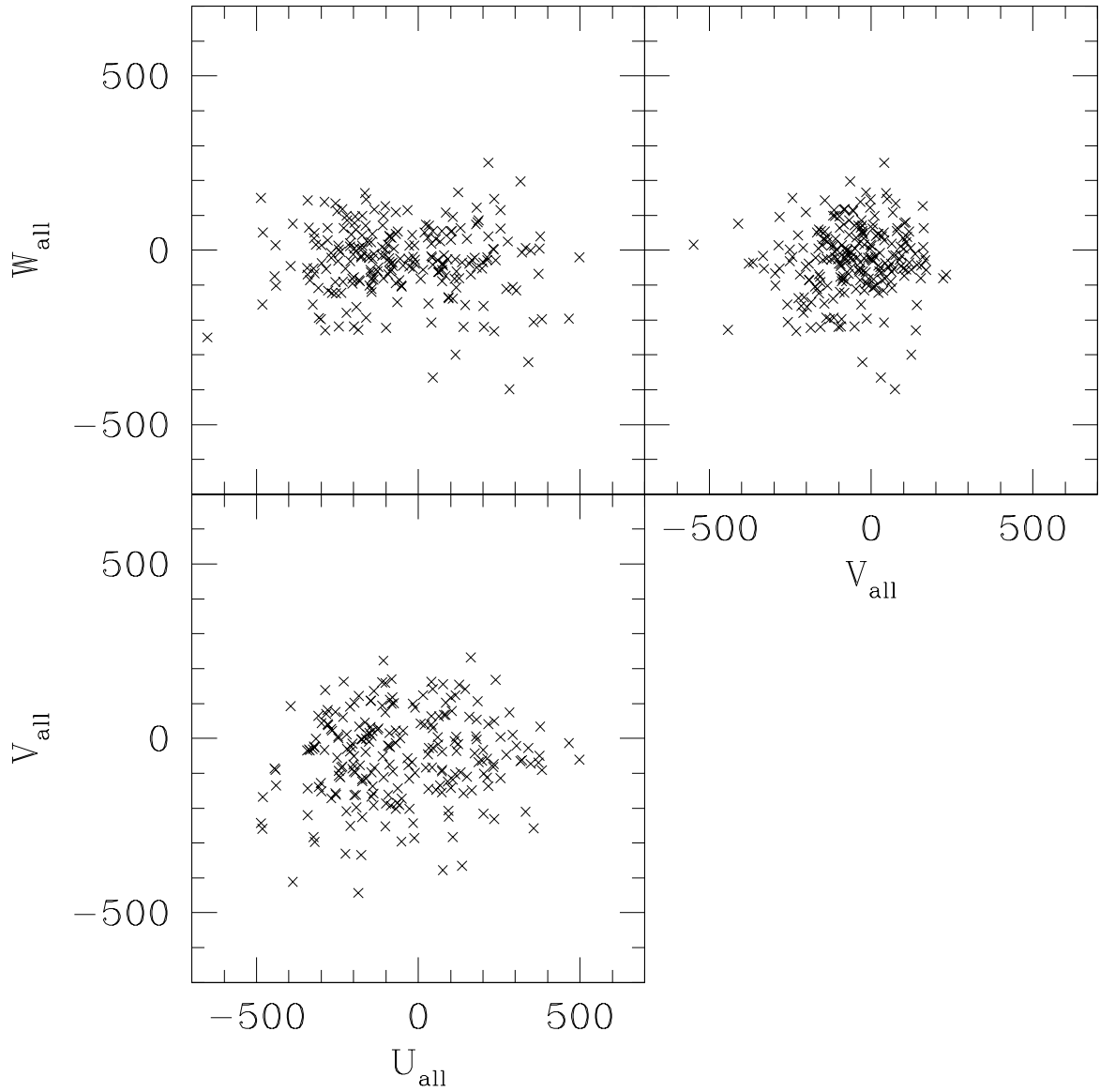}
\caption{$UVW$ velocities (in Galactocentric frame) of 295
stars selected from rNLTT using the same $1<\eta<4.15$ criterion adopted
by \citet{gould03a}.  There is no significant contamination by thick-disk
stars.  Kindly provided by J.\ Marshall in advance of publication.}
\label{fig:uvwallzoom}
\end{figure}

J.\ Marshall (private communication 2007) has also obtained 
radial velocities for 295 of the stars in her sample, which
(together with her excellent photometry and the rNLTT proper motions)
permit her to make 3-dimensional velocity estimates for each star.
The results are shown in Figure \ref{fig:uvwallzoom}.  The
distribution is not significantly contaminated by thick-disk stars,
which would appear as an overdensity centered at 
$(U,V,W)\sim (0,190,0)\,\kms$.

\clearpage

\section{Metallicity Distribution vs.\ Kinematic Cuts
\label{sec:metallicity}}

\begin{figure}
\includegraphics[angle=0,width=3.25in]{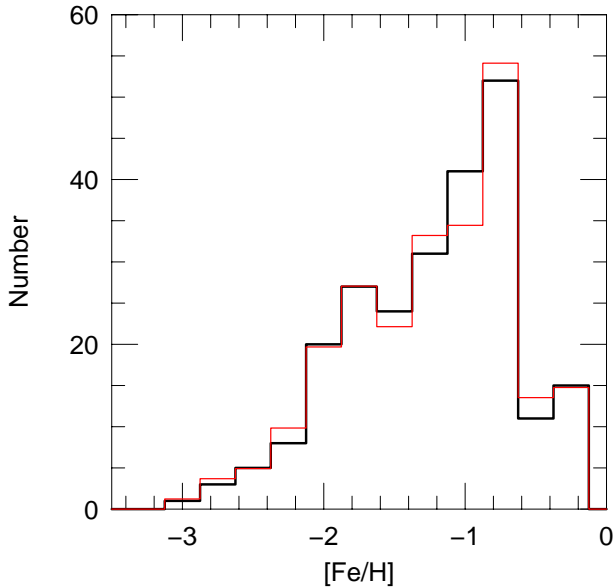}
\caption{Metallicity distribution for 239 
stars selected from rNLTT using the same $1<\eta<4.15$ criterion adopted
by \citet{gould03a} ({\it black}) and for a ``kinematically secure halo''
subsample of 194 stars ({\it red}).  The latter histogram is scaled
up by 239/194 so that the sum of the bins is the same.  The
two histograms are essentially identical, while the analogous Figure 2
of \citet{ryan91} shows that their NLTT-derived sample was heavily
contaminated by thick-disk stars.
Kindly provided by J.\ Marshall in advance of publication.}
\label{fig:vfeh}
\end{figure}

J.\ Marshall (private communication 2007) was able to make
preliminary metallicity measurements for 239 of the 295 halo candidates
mentioned in \S~\ref{sec:velocity}.  These permit a direct comparison
with the test performed by \citet{ryan91}, by which they showed that
the halo candidates that they had selected from NLTT were heavily contaminated
by thick-disk stars.  Figure \ref{fig:vfeh} is the direct analog
of Figure 2 from \citet{ryan91}.  It shows the metallicity for all
239 stars and also for a subsample of 194 ``kinematically secure halo stars''
defined as the union of stars with either $v_\perp>220\,\kms$ or
velocity in the rotation direction $V<-220\,\kms$.  The first point to note
is that 82\% of the stars are ``secure halo'' by this definition.  The
second point is that the metallicity distributions of the two samples
are essentially identical.  By contrast, \citet{ryan91} found a huge
tail of higher-metallicity stars in their full sample, which disappeared
when they implemented the same ``secure halo star'' cut.
This brings the argument full circle.

\section{Conclusion
\label{sec:conclusion}}

Contamination of the original \citet{gould03a} halo sample by thick-disk
stars is likely to be of order 2\%.  This implies that the effect
of streams is not ``washed out'' by thick-disk contamination.
Hence, the limits on granularity found by \citet{gould03b} in this
sample imply corresponding limits on streams in the Galactic halo.
In particular, at 95\% confidence, no more than 5\% of local halo
stars (within about $300\,$pc) are in any one coherent stream.

\acknowledgments
I thank Juna Kollmeier for suggesting a number of improvements
to the original manuscript.
I am grateful to Jennifer Marshall for making available her
plots of 3-D velocity (Fig.\ \ref{fig:uvwallzoom}) 
and metallicity distribution (Fig.\ \ref{fig:vfeh}) in advance of
publication.
This work was supported in part by
grant AST-042758 from the NSF.

\end{document}